\documentclass[12pt]{article}
\usepackage[centertags]{amsmath}
\usepackage{amssymb}
\usepackage{amsthm}
\usepackage{enumerate}
\usepackage{url}
\usepackage{graphicx}

\usepackage{graphicx}
\usepackage[all, knot]{xy}
\usepackage{amsfonts,amssymb}
\usepackage{latexsym}
\usepackage{amsmath}
\usepackage[all]{xy}
\usepackage{stmaryrd}

\usepackage[english]{babel}

\usepackage[small,nohug,heads=vee]{diagrams}
\diagramstyle[labelstyle=\scriptstyle]

\usepackage{color}
\usepackage[normalem]{ulem}
\usepackage{url}

\makeatletter
  \def\tagform@#1{\maketag@@@{(#1)\@@italiccorr}}
\makeatother

\newcommand {\cE}{\mbox{${\mathcal E}$}}

\newcommand {\cM}{\mbox{${\mathcal M}$}}

\newcommand{\barM}{\bar{M}}
\newcommand{\R}{\mathbb{R}}

\newcommand{\MM}{\mathbb{M}}
\newcommand{\NN}{\mathbb{N}}

\begin{document}

\title{Beyond the Space-Time Boundary}
\author{Michael Heller \\
Copernicus Center for Interdisciplinary Studies\\
ul. Szczepa\'nska 1/5, 31-011 Cracow, Poland\\
\and 
Jerzy Kr\'ol \\
Institute of Physics, University of Silesia, \\
ul. Uniwersytecka 
4, 40-007 Katowice, Poland}

\date{\today}
\maketitle

\begin{abstract}
In General Relativity a space-time $M$ is regarded singular if there is an obstacle that prevents an incomplete curve in $M$ to be continued. Usually, such a space-time is completed to form $\bar{M} = M \cup \partial M$ where $\partial M$ is a singular boundary of $M$. The standard geometric tools on $M$ do not allow ``to cross the boundary''. However, the so-called Synthetic Differential Geometry (SDG), a categorical version of standard differential geometry based on intuitionistic logic, has at its disposal tools permitting doing so. Owing to the existence of infinitesimals one is able to penetrate ``germs of manifolds'' that are not visible from the standard perspective. We present a simple model showing what happens ``beyond the boundary'' and when the singularity is finally attained. The model is purely mathematical and is mathematically rigorous but it does not pretend to refer to the physical universe.

\end{abstract}

\section{Introduction}
There is a general agreement among specialists that ``to say that a space-time is singular means that there is some positive obstacle that prevents an incomplete curve continuing'' \cite{Clarke}\footnote{The present paper is based on a talk delivered at the conference ``Category Theory in Physics, Mathematics and Philosophy`` held at the Warsaw University of Technology, 16-17 November 2017.}. Very roughly speaking there are two kinds of such obstacles: (1) some magnitudes, such as curvature or some scalars constructed from it, become unbounded along a timelike curve before it ends, or (2) a pathological behaviour of the differential structure of space-time prevents a timelike curve from being prolonged. The book by Clarke \cite{Clarke} is almost exclusively devoted to make precise and understand the second of these obstructions. In the present paper, we continue this line of research but we essentially change the method of investigation. In the meantime (after the publication of Clarke's book) a new approach in mathematics has matured that not so much solves but rather circumvents many problems related to differentiability. We have in mind the so-called Synthetic Differential Geometry (SDG), an extension of the usual differential geometry, based on category theory (fundamental monographs are \cite{Kock2006,Kock09,MoerRey}). The fact that this approach enforces the employment of intuitionistic logic results into enriching the real line $\R $ with various kinds of infinitesimals. We may imagine that they constitute the entire world inside every point of $\R $, a sort of a fiber over $x \in \R $. Owing to the existence of infinitesimals differentiation becomes a purely algebraic operation and every function is differentiable as many times as required. This creates a unique opportunity for facing the problem of space-time prolongations and singularities in General Relativity (GR).

We approach this problem along the following lines. In section 2, we give necessary preliminaries: we define these kinds of infinitesimals that are used in the sequel, we define the ``$k$th order neighbouring relation'', and in terms of it the concept of monad -- a kind of minimal portion of space.
We also quote some of its properties.

In section 3, we present an ``infinitesimal version'' of the usual differential manifold concept, called formal $n$-dimensional manifold. In such a formal manifold, monads are domains of local maps.

In section 4, we present our model which shows, by incorporating the machinery sketched in the previous sections, what happens ``beyond'' the singular boundary $\partial M$ of space-time $M$. We assume that this boundary contains a strong curvature singularity such as the Big Bang type of singularity.

We add two appendices. The first appendix gives a more detailed mathematical description of how the transition from $M$ to $\partial M$ could look like. The second appendix presents a concept that could be useful in studying a dynamics of monads.

Although the model attempts to imitate, for pedagogical reasons, the standard evolution of the Friedman-Lema\^{i}tre-Robertson-Walker (FLRW) cosmological model, it does not pretend to describe the actual evolution of the universe. At its present stage of development it is nothing more but a purely mathematical exercise.

\section{Preliminaries}
In SDG, one considers various kinds of infinitesimals\footnote{In general, infinitesimal objects are identified with spectra of a Weil algebras.}. Let us denote by $R $ the real line $\R $ enriched by infinitesimals. In the present study, we focus on the following ones
\[
D = \{x \in R|x^2=0 \},
\]
\[
D_k=\{ x\in R|x^{k+1}=0 \}, \; k=1,2,3,..., 
\] 
\[ 
D(n)=\{ (x_1,...,x_n)\in R^n|x_ix_j=0,\; \forall i,j=1,2,3,...,n \}, 
\] 
\[
 D_k(n)=\{(x_1,...,x_n)\in R^n|{\rm \, the\, product\, of \, any}\,  
k+1\, {\rm of} \, x_i\, {\rm is}\, 0 \}, 
\] 
and finally, 
\[
 (D_{\infty})^n=\bigcup_{k=1}^{\infty}D_k(n).
\]

Analogously, we can define $D(V)$ and $D_k(V)$ for any finite dimensional vector space $V$.

We can imagine infinitesimals as internal ``degrees of freedom'' of a single point of $\R $.

In what follows, our important tool is the ``$k$th order neighbouring relation'', defined as
\[
u \sim_k v \Leftrightarrow  u - v \in D_k(V).
\]
This relation is reflexive and symmetric, but it is not transitive; instead we have
\[
(u\sim_kv \wedge v\sim_l w) \Rightarrow (u \sim_{k+l} w).
\]

We assume that everything in this section happens in a ``suitable'' category \cE . ``Suitable'' means a category equipped, among others, with a commutative ring object $R$, usually a topos. We further assume that the category $\MM $ of manifolds (and smooth functions) can be regarded as a subcategory of \cE . Let $M$ be an $n$-dimensional (formal) manifold considered as an object of $\cE $. We are interested in the ``smallest neighbourhoods'' of $M$. A good tool to investigate such neighbourhoods is the neighbourhood relation $\sim_k$ but it should first be generalised to the manifold context. Let $x, y \in M$ and $k$ be a non-negative natural number; the relation $x \sim_k y$ holds iff there exists a coordinate chart $f: U \to M$ such that $U\subseteq R^n$ is open, and in $U$ we have $f(x) \sim_k f(y)$. If $k=1$, we simply write $x\sim y$.

Let $M$ be a manifold and $x, y, z \in M$. The neighbouring relation $\sim $ satisfies the  following conditions
\begin{enumerate}
\item
$x \sim_0y$ iff $x=y$ (reflexivity).
\item
$x\sim_ky$ implies $y\sim_lx$ if $k\leq l$ (symmetry),
\item
$x\sim_ky$ and $y\sim_lz$ implies $x\sim_{k+l}z$ (quasi-triangle formula).
\end{enumerate}

Since these conditions are akin to the usual concept of distance, we can define a ``quasi-distance'' function in the following way
\[
\mathrm{dist}(x,y) \leq k \;\; \mathrm{if} \;\; x\sim_ky.
\]
Since $D_k(n) \subseteq D_l(n)$ if $k \leq l$, the function ``dist'' determines a ``size'' of an object $D_k(n)$. Let us also notice that this quasi-metric is ``quantised'' (discrete) since it has its values in $\NN $.

Now, we define a few key concepts for our further considerations.

Let $x \in M$. The $k$-monad around $x$ is defined to be
\[
\cM_k(x) := \{y\in M| x\sim_k y\} \subseteq M.
\]
If $k=1$, we write $\cM(x)$. We also assume that $\cM_{\infty }$ makes sense. We obviously have $y\in \cM_k(x) \Leftrightarrow x\in \cM_k(y)$. If $f: M\to N$ is a map between manifolds $M$ and $N$ then $x\sim_k y$ implies $f(x) \sim_k f(y)$, since in SDG every map $f:D_k\to R$ such that $0\mapsto 0$, factorizes through $D_k$ \cite[Cor. 6.2]{Kock2006}.

The ``$k$th neighbourhood of the diagonal'', $M_{(k)} \subseteq M \times M$, is defined to be
\[
M_{(k)} := \{(x,y) \in M \times M|x \sim_k y\}.
\]
If $V$ is an $n$-dimensional vector space, there is a canonical isomorphism
\[
M_{(k)}\cong M \times D_k(V)
\]
given by
\[
(x,y) \mapsto (x, y-x),
\]
and consequently there is an isomorphism between $\cM_k(x)$ and $D_k(V)$ \cite[p. 39]{Kock09}. Let us also notice that the quasi-metric dist$(d_1, d_2) \leq k$ introduces a partial order in $\cM_{\infty }(x)$ (by inclusions).

\section{Manifold in the Smallest}
The existence of infinitesimals essentially enriches the structure of differential manifolds. It enables the following definition (\cite{Kock80}, see also \cite[pp.68-71]{Kock09}). An object $M$ in the category $\cE $ is said to be a $k$-formal $n$-dimensional manifold if, for each $x \in M$, there exists a monad $\cM_k(x)$, isomorphic to $D_k(n)$, and   a map $f: \cM_k(x) \to M$. A bijective map $D_k(n) \to \cM_k(x)$ onto a monad around $x$, mapping $0$ to $x$, is said to be a $k$-frame at $x$. $k$ can assume the value $\infty $.  If $\cM_{\infty }(x)$, we speak of a formal $n$-dimensional manifold (without specifying $k$).

It can be easily seen that $R^n$, for every $n$, is a formal $n$-dimensional manifold, and the monad $\cM_\infty(v)$ around $v \in R^n$ is $\cM(v) = v + D_{\infty }(n)$.

We have the following natural, but important, results.

If $M$ and $N$ are formal manifolds of dimensions $m$ and $n$, respectively, then $M \times N$ is an $(m+n)$-dimensional formal manifold; and the monad around $(x, y) \in M \times N$ is $\cM (x) \times \cM (y)$ which is isomorphic to $D_{\infty }(m+n)$.

If $M$ is a formal $n$ dimensional manifold, then its tangent bundle $M^D$ is a $(m+n)$-dimensional formal manifold.

\section{A Model}
Let us consider a singular space-time; it is singular in the sense that it contains at least one incomplete curve that cannot be continued in any extension of this space-time. Regular part of this space-time forms a differential manifold $M$. We define the completion of $M$ as $\barM = M \cup \partial M$ and call $\partial M$ the singular boundary of $M$. We assume that this boundary is attainable from $M$, i.e. that $M$ is open and dense in $\barM$. Details of this construction are of no importance for our further analysis (there are known several proposals, such as $g$-boundary, $b$-boundary, causal boundary, and others).

Singular boundary can contain, besides endpoints of inextendible curves that cannot be continued in any extension of space-time, also endpoints of curves that can be continued in some of its extensions, and ``points at infinity''. In what follows, we assume, for simplicity, that $\partial M$ contains only endpoints of inextendible curves that cannot be continued in any extension of space-time.

For the sake of concreteness let us think about the FLRW space-time with the Big Bang singularity (strong curvature singularity) in the beginning (the central Schwarzschild singularity would also fit the picture), and let us contemplate the evolution of the universe back in time. Everything happens according to the standard cosmological model. The universe shrinks, subsequent cosmic eras succeed each other. Finally, the contraction attains the state in which differential properties of space time break down completely. The universe leaves the ``manifold region'' $M$ end enters its boundary $\partial M$. This means that the standard smooth manifold description breaks down, and we assume that at this stage the category $\cE $ takes over (see appendix 1). The contraction has reached  such a degree that infinitesimals enter into play. We thus can employ methods of SDG to gain an insight into what is going on. General picture that emerges is the following.

After crossing $\partial M$, domains $U$ of local charts of the manifold $M$, $U \to M$, become infinitesimal, and the manifold becomes 4-dimensional formal manifold as defined in section 3. Local charts are now of the form $\cM_{\infty }(x) \to M$ (the fact that we use the same letter for denoting the space-time manifold and the formal manifold should not lead to misunderstandings). But the universe continues shrinking, and finally its size reduces to a single monad $\cM_{\infty }(x_0)$. The contraction goes on, but now only in the sense of the metric dist$(d_1, d_2) \leq k, \; d_1, d_2 \in \cM_{\infty }(x_0)$. This means that the differentiability properties are lower and lower, and we obtain a decreasing sequence
\[
\cM_{\infty }(x_0), \ldots , \cM_k(x_0), \cM_{k-1}(x_0), \cM_{k-2}(x_0), \ldots
\].

Finally, when the contraction produces $\cM_0(x_0)$, the process comes to a halt, since  $\cM_0(x_0) = \{y \in R| x_0\sim_0 y\}$ which, in turn gives $x_0=y$, and all quasi-distances $\mathrm{dist}(x_0,y)$ reduce to zero.

It is instructive to follow the entire process starting from the zero-state. It seems natural to regard increasing sequence of $k$s as a sort of quantised time. If this seems to you too farfetched, you can treat the ``time'' as just a name of a parameter. However, it is important to notice that $k$, in fact, means the ``degree of differentiability'', the place at which the Taylor expansion truncates (all higher order terms vanish). Each subsequent instant of this time improves differential properties of the process.

It is astonishing that the transition from $k=0$ to $k=1$ is so exuberantly reach. Having at our disposal the very first degree of differentiability, we can do large parts of affine geometry, affine connection included, combinatorial differential forms, tangent bundle and lot of differential geometry (in fact, a substantial part of Kock's seminal SDG monograph \cite{Kock09} is limited to explore the geometry the first order neighbourhoods).

Of course, when we jump to $k = 2$, differentially-geometric properties substantially improve. Some aspects of metric geometry came into force \cite[chapter 8]{Kock09}.

For doing $k$-jet theory, we evidently need a sufficiently high $k$. 

Finally, when $k \to \infty $, we end up in $\cM_{\infty }(x_0)$ and we have the full differentiability. As the universe expands, it goes through the phase of a formal manifold, and when infinitesimals cease to play any role (because of the expansion), the standard smooth manifold regime takes over. In the language of space-time boundary, this means that the universe goes from $\partial M  $ to $M$. It is here that we should place the transition from the category $\cE $ to the category SET, and possibly identify this with what physicists call Planck's threshold.

\setcounter{section}{0}
\section{Appendix: Through the Boundary}
In this appendix, we give a short mathematical description of how the transition from $M$ to $\partial M$ could look like. The crucial point is that if we go from space-time $M$ to its singular boundary $\partial M$, we must switch from the category SET of sets and maps between sets as morphisms to a suitable category, to which we have assigned the symbol $\cE $, the internal logic of which is intuitionistic that enables infinitesimals to appear.

Space-time $M$ is supposed, as always, to be a smooth paracompact manifold. First, we move from its description in terms of maps and atlases to the functional description in terms of the algebra $C^{\infty }(M)$ of smooth functions on $M$. This is possible owing to the generalized Gelfand-Naimark theorem which asserts that the category of locally compact Hausdorff spaces and proper continuous maps is anti-equivalent to the category of commutative $C^*$-algebras and nondegenerate morphisms \cite{Landsman}. We remind that a map $\phi: X \to Y$ between locally compact Hausdorff spaces is said to be  proper if, for any compact $K \subset Y$, $\phi^{-1}(K)$ is compact in $X$. And a morphism $\psi : A \to B$ between $C^*$-algebras is said to be nondegenerate, if the linear span of all expressions of the form $\psi (a)b$, $a \in A, \,b \in B$, is dense in $B$. Two categories $G$ and $H$ are said to be anti-equivalent (or dual) if there exist contravariant functors $\alpha : G \to H$ and $\beta : G\to H$ such that $\alpha \circ \beta $ and $\beta \circ \alpha $ are naturally isomorphic to $id_H$ and $id_G$, respectively.

We now look at $C^{\infty }(M)$ from another perspective. A smooth algebra (or $C^{\infty }$-ring) is an algebra $A$ over $\R $ for which the product $ \cdot : \R \times \R \to \R$ lifts to the algebra product $A \times A \to A$, and also every smooth map $f: \R^n \times \R^m$ lifts to a map $A(f): A^n \to A^m$, in such a way that projections, identities and compositions are preserved. Formally, such a smooth algebra is a functor from the category Cart of Cartesian spaces to the category SET,
\[
A: \mathrm{Cart} \to \mathrm{SET},
\]
that preserves finite products. The category of such functors as objects and natural transformations between them as morphisms is denoted by $C^{\infty }$-Alg \cite[pp. 15-16]{MoerRey}.

If $M \in \MM$ is a smooth manifold then the functor
\[
C^{\infty }(M) = \mathrm{Hom}_{\MM }(M, -)
\]
is a smooth algebra $C^{\infty }(M)$. Considering an object (here a smooth manifold) in another category may not be a superficial change. Changing a categorical context can not only provide new tools of investigation, but can also affect properties of the object itself. Let us pursue this line of research.

A smooth algebra $A$ is said to be  finitely generated if it is of the form $C^{\infty }(\R^n)/I$, for $n \in \NN $ and an ideal I; if additionally $I$ is finitely generated, $A$ is said to be finitely presented. For every smooth manifold $M$, the smooth algebra $C^{\infty }(M)$ is finitely presented \cite[p. 25]{MoerRey}.

Finitely generated smooth algebras and $C^{\infty }$-homomorphisms between them as morphisms form a category denoted by FGAlg (see \cite[p. 21]{MoerRey}). We define the category LOC (of loci) as the opposite category with respect to FGAlg. The objects of LOC are the same as those of FGAlg (if $A \in $ FGAlg, we shall write $lA $ when $A$ is considered as an object of LOC), but morphisms are reversed. Since the functor from the manifold category to $C^{\infty }$-Alg, given by $M \mapsto C^{\infty }(M)$, is contravariant, to obtain the correct variance, we switch to the category LOC. And indeed, we have the covariant functor $s: \MM \to $LOC, given by $M \mapsto lC^{\infty }(M)$ which, importantly, is full and faithful.

This switching the categorical environment produces radical changes. A new element that appears in this context is the so-called Weil algebra. It is a finite dimensional $\R $-algebra $W$ having a maximal ideal $I$ such that $W/I \simeq \R$ with $I^n = 0$ for some $n \in \NN $. It has a unique smooth algebra structure and is finitely presented. Objects of LOC (smooth loci) corresponding to Weil algebras are infinitesimal spaces. In this way, infinitesimals appear in our model. We immediately have: $C^{\infty }(\R ) \in $\, LOC is a real line enriched with infinitesimals which we denote by $R$.

We now have a functor $s: \MM \to $\, LOC, given by $ M \mapsto lC^{\infty }(M)$ which is again full and faithful. However $s(M)$ is much richer than $M$. It contains infinitesimal portions of a manifold or spaces which can be called ``germs of a manifold''. To define them, let us notice that if $lA = C^{\infty }(\R^n)/I$ then $p: 1 \to lC^{\infty }(\R^0)$, where $1 = lC^{\infty}(\R^0)$, is a point in $lA$. Let now $p \in \R$. We define the germ $C^{\infty }_p(\R )$ of $R$ at $p$ to be $\bigcap \{s(U)| p \in U \mathrm{open \; in}\; \R \}$. And analogously for $p \in \R^n$. We are now ready to define the germ of $lA $ at $p \in \R^n$ as $lC^{\infty }_p(\R^n) \cap lA$. This of course remains valid if $lA = s(M)$.

Let us consider a function $f: M \to \R $, and the germ $f_p$ of this function at $p \in M$ (in the usual sense). This germ can now be identified with the restriction of the function $f$ to the germ of the manifold $M$ at $p$ (for details see \cite[p. 64]{MoerRey}).

We now are ``well inside'' the boundary $\partial M$. As the universe continues shrinking, domains of local maps become infinitesimals, and the usual manifold is replaced by a formal manifold. Finally, when shrinking still progresses, everything is reduced to a single monad $\cM_{\infty }(x_0)$. 

Usually, to improve geometric properties of a given model, one once again changes the category LOC to the category $\mathrm{SET}^{\mathrm{LOC}^{op}}$  of presheaves on LOC or to some of its subcategories. In our case, this does not seem indispensable since LOC has good properties if limited to ``sufficiently small'' spaces \cite[p. 71]{MoerRey}.

\section{Appendix: Non-Holonomous Monads}
In studying interactions between different levels of differentiability in a monad the following concept is useful.

Let $M$ be a (formal) manifold, and let us consider a sequence $k_1, \dots , k_r$ of non-negative integers. Let us also  consider the set $M_{(k_1, \dots , k_r)} \subseteq M^{r+1}$, the elements of which are $r+1$-tuples
\[
X = (x_0, k_1, \dots , k_r)
\]
such that
\[
x_0 \sim_{k_1} x_1 \sim_{k_2} \ldots \sim_r x_.
\]
If $r=1$, we recover $M_{k_1}$. Assigning to such an $r+1$-tuple its first element $x_0$ (when $x_0$ varies over $M$), we obtain a bundle $M_{(k_1, \dots , k_r)} \to M$ over $M$. The fiber over $x=x_0$ of this bundle is called the non-holonomous $(k_1, \ldots , k_r)$-monad around $x$ and is denoted by $\cM_{k_1, \ldots , k_r}(x)$. There exists a map
\[
\cM_{(k_1, \ldots , k_r)}(x) \to \cM_{(k_1 + \ldots + k_r)}
\]
transforming non-holonomous monads into ordinary (holonomous) ones (for more see \cite[pp. 86-88]{Kock09}).

\end{document}